\documentclass[aps,prl,twocolumn,superscriptaddress]{revtex4-2}
\usepackage{newtxtext}
\usepackage{lipsum}

\usepackage{amsmath}
\usepackage{amssymb}
\usepackage{mathtools}
\usepackage{graphicx}
\usepackage{xcolor}
\usepackage{bm}
\usepackage{mathrsfs}
\usepackage{colortbl}

\definecolor{myred}{HTML}{E1558A}
\definecolor{myblue}{HTML}{318294}
\definecolor{myskyblue}{HTML}{3FC3E0}
\definecolor{addition}{HTML}{EAAD00}
\definecolor{reduction}{HTML}{124096}

\newcommand{\curr}{\bm{\mathsf{J}}}
\newcommand{\revcurr}{\curr^{\mathrm{rev}}}
\newcommand{\irrcurr}{\curr^{\mathrm{irr}}}
\newcommand{\force}{\bm{\mathsf{F}}}
\newcommand{\ccdot}{\!:\!}
\newcommand{\cons}{\bm{\Pi}_\rho}

\usepackage[hidelinks]{hyperref}
\hypersetup{
  colorlinks   = true, 
  urlcolor     = blue, 
  linkcolor    = myred, 
  citecolor   = myblue 
}

\newcommand{\tr}{\mathop{\mathrm{tr}}\nolimits}
\begin{document}

\title{Two applications of stochastic thermodynamics to hydrodynamics}
\author{Kohei Yoshimura}
\email{kyoshimura@ubi.s.u-tokyo.ac.jp}
\affiliation{Department of Physics, The University of Tokyo, 7-3-1 Hongo, Bunkyo-ku, Tokyo 113-0033, Japan}
\author{Sosuke Ito}
\affiliation{Department of Physics, The University of Tokyo, 7-3-1 Hongo, Bunkyo-ku, Tokyo 113-0033, Japan}
\affiliation{Universal Biology Institute, The University of Tokyo, 7-3-1 Hongo, Bunkyo-ku, Tokyo 113-0033, Japan}
\date{\today}

\begin{abstract}
    Recently, the theoretical framework of stochastic thermodynamics has been revealed to be useful for macroscopic systems. However, despite its conceptual and practical importance, the connection to hydrodynamics has yet to be explored. 
    In this Letter, we reformulate the thermodynamics of compressible and incompressible Newtonian fluids so that it becomes comparable to stochastic thermodynamics and unveil their connections; we obtain the housekeeping--excess decomposition of entropy production rate (EPR) for hydrodynamic systems and find a lower bound on EPR given by relative fluctuation similar to the thermodynamic uncertainty relation. 
    These results not only prove the universality of stochastic thermodynamics but also suggest the potential extensibility of the thermodynamic theory of hydrodynamic systems. 
\end{abstract}

\maketitle

\begin{figure}[t]
    \centering
    \includegraphics[width=\linewidth]{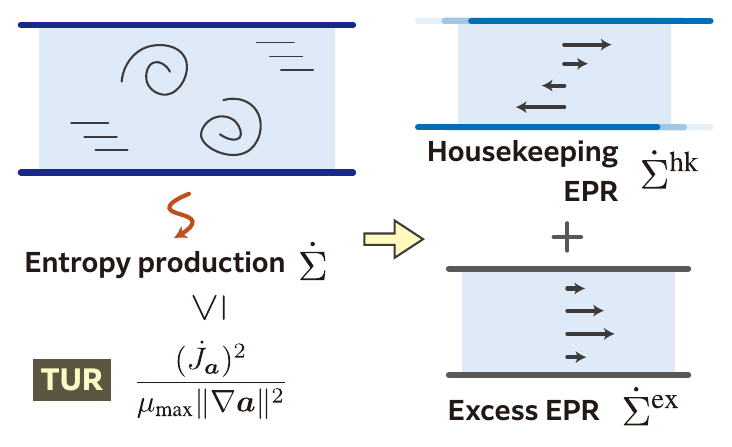}
    \caption{In this letter, we employ techniques of stochastic thermodynamics to derive housekeeping--excess decomposition of the EPR and a TUR. Respectively, the housekeeping and excess EPRs capture dissipation due to external driving that keeps the system out of equilibrium and dissipation arising from the remaining transient part that behaves as if the system were not driven. We also show that they can be defined geometrically and that the excess EPR can be interpreted as the minimum dissipation. The TUR is given by any appropriate vector field and is inversely proportional to the viscosity coefficient and the spatial fluctuation of the field. }
    \label{fig:concept}
\end{figure}

\textit{Introduction}.---%
The second law of thermodynamics is the most fundamental, universal restriction on what physical systems can do. 
This century, its detailed character has been revealed by stochastic thermodynamics in thermally fluctuating nonequilibrium systems, which can be classical or quantum, relying on entropy production as a critical quantity~\cite{seifert2012stochastic,landi2021irreversible}. 
Despite the significant development of our knowledge of entropy production and the second law in such systems~\cite{jarzynski1997nonequilibrium,barato2015thermodynamic,shiraishi2018speed}, application of the developed techniques to other types of systems is not well examined, except for deterministic chemical systems~\cite{rao2016nonequilibrium,yoshimura2021thermodynamic}. 

Deterministic hydrodynamic systems described by the Navier--Stokes equation are among the least investigated subjects. 
Thermodynamics of such systems was once intensively studied in the last century~\cite{de1962non}, but it has yet to be considered from the viewpoint of stochastic thermodynamics.
Nonetheless, a universal understanding of hydrodynamic systems as provided by thermodynamics is no less valuable than that of thermally fluctuating or chemical systems because the Navier--Stokes equation governs many phenomena ranging from the motion of tiny cells~\cite{lauga2009hydrodynamics,marchetti2013hydrodynamics} to daily water usage and industrial water management~\cite{chadwick2021hydraulics}. 
In particular, entropy production has recently been attracting attention due to its practical importance in evaluating the performance of hydraulic machinery~\cite{zhou2022application}. 
However, modern knowledge of thermodynamics that stochastic thermodynamics has yielded has been far from utilized for those systems. 

In this Letter, we develop two ways to apply stochastic thermodynamics to hydrodynamics, summarized in Fig.~\ref{fig:concept}: First, we define a decomposition of entropy production rate (EPR), which is called the housekeeping--excess decomposition and has been studied in stochastic thermodynamics for the past quarter century~\cite{oono1998steady,hatano2001steady,esposito2010three,rao2016nonequilibrium,ge2016nonequilibrium,maes2014nonequilibrium,dechant2022geometric1,dechant2022geometric2,yoshimura2023housekeeping,kolchinsky2022information,kobayashi2022hessian}.
It was first proposed to recover the second law of thermodynamics, which becomes futile in quasi-static processes between nonequilibrium steady states~\cite{oono1998steady,hatano2001steady,esposito2010three,rao2016nonequilibrium,ge2016nonequilibrium,maes2014nonequilibrium}.
Now, it is known that it can provide tighter thermodynamic trade-off relations reflecting the nonequilibrium situation of the system appropriately~\cite{shiraishi2018speed,tuan2020unified,dechant2022geometric1,dechant2022geometric2,yoshimura2023housekeeping,kolchinsky2022information}. 
Second, we derive a lower bound on the EPR that resembles what is known as the thermodynamic uncertainty relations (TURs)~\cite{barato2015thermodynamic,pietzonka2016universal,dechant2018multidimensional,liu2020thermodynamic,horowitz2020thermodynamic,yoshimura2021thermodynamic,li2019quantifying,otsubo2020estimating,manikandan2020inferring}, one of the most privileged thermodynamic trade-off relations.
The TURs have two aspects: They indicate trade-off relations between dissipation and fluctuations~\cite{barato2015thermodynamic,pietzonka2016universal} or provide EPR estimators~\cite{li2019quantifying,otsubo2020estimating,manikandan2020inferring}. 
Our TUR is of the first kind; it gives a lower bound with a changing rate of a vector field and two fluctuations, one intrinsic in the system and the other the vector field exhibits. 
In the two applications, the key is the geometric expression of EPR and the definition of conservative systems. 

\textit{Preliminaries}.---%
We consider a compressible Newtonian fluid in a $n$-dimensional connected region $\Omega$ with boundary $\partial\Omega$. 
We assume the system is locally equilibrated so that we can define thermodynamic quantities locally~\cite{de1962non}. 
The temperature is supposed to be homogeneous, so the dissipation due to heat flow is absent. We especially set the temperature to the unity. 
The extension to the incompressible systems is discussed in Supplementary Material~\cite{smArxiv}.

The state of a system is designated by the density field $\rho(\bm{x})$ and the velocity field $\bm{v}(\bm{x})=(v_i(\bm{x}))_{i=1}^n$ at each point $\bm{x}\in\Omega$. 
When considering the boundary conditions of velocity fields, we focus on the values rather than the derivatives. 
The dynamics are described by the continuity equations of density $\rho$ and momentum $\rho\bm{v}$ as~\cite{batchelor1967introduction}
\begin{align}
    \partial_t\rho=-\nabla\cdot(\rho\bm{v}),\quad
    \partial_t(\rho\bm{v})
    =-\nabla\cdot\curr.
\end{align}
Here, $\curr$ denotes the momentum current, decomposed into the reversible and irreversible currents as $\curr=\revcurr+\irrcurr$. 
The reversible part is given as $\revcurr=\rho\bm{v}\otimes\bm{v}+p\bm{\mathsf{I}}$ with pressure $p$ and the identity matrix $\bm{\mathsf{I}}$. Note $\bm{v}\otimes\bm{v}$ is a matrix whose $(i,j)$-element is $v_iv_j$. 
The irreversible current expresses the momentum transfer via viscous stress. For the Newtonian fluids, it is provided as $\irrcurr=-(\zeta-\frac{2}{n}\mu)(\nabla\cdot\bm{v})\bm{\mathsf{I}}-\mu[\nabla\bm{v}+(\nabla\bm{v})^\mathsf{T}]$ with the volume and shear viscosities $\zeta$ and $\mu$, which can be dependent on the density. 
Note that $\nabla\bm{v}$ is the matrix with elements $\partial_iv_j$ and the superscript ${}^\mathsf{T}$ represents transposition. 
For later convenience, we define $\lambda\coloneqq\zeta-\frac{2}{n}\mu$. 
In addition, we write the symmetrized gradient $[\nabla\bm{v}+(\nabla\bm{v})^\mathsf{T}]/2$ as $\nabla^\mathsf{S}\bm{v}$. 
As a result, the irreversible current can be rewritten as $\irrcurr=-\lambda(\nabla\cdot\bm{v})\bm{\mathsf{I}}-2\mu\nabla^\mathsf{S}\bm{v}$. 

The continuity equation turns out to be the renowned Navier--Stokes equation if we rewrite it as
\begin{align}
    \rho\frac{D\bm{v}}{Dt}=\nabla\cdot\bm{\sigma},
\end{align}
where $D/Dt\coloneqq\partial_t+\bm{v}\cdot\nabla$ and $\bm{\sigma}=-p\bm{\mathsf{I}}-\irrcurr$. 
We can define the pathline $\bm{\phi}_t(\bm{x})$ as the solution of $\partial_t\bm{\phi}_t(\bm{x})=\bm{v}(\bm{\phi}_t(\bm{x}),t)$ with the initial condition $\bm{\phi}_0(\bm{x})=\bm{x}$. 
It gives the trajectory of a particle starting from $\bm{x}$ in the fluid. 

The local equilibrium assumption allows us to discuss the entropy production rate (EPR). 
In general, EPR is given as the product between the irreversible current and the thermodynamic force that induces the current~\cite{de1962non}. 
Not following the traditional way, we regard the symmetrized gradient of a velocity field as a thermodynamic force. 
That is, the thermodynamic force $\force$ of a system $(\rho,\bm{v})$ is defined as $\force=-\nabla^\mathsf{S}\bm{v}$. 
The thermodynamic forces are connected to the irreversible currents through the constitutive equation $\irrcurr=\cons(\force)$, the form of which depends on the system. Here, $\rho$ indicates its dependence on the density field. 
Then, the EPR can be expressed as $\dot{\Sigma}=\int_\Omega\cons(\force)\ccdot\force\,dx$, where colon $:$ denotes the inner product between two matrices; $A\ccdot B\coloneqq\sum_{i,j}A_{ij}B_{ij}$.
Inspired by the expression, we define an inner product for symmetric tensor fields as
\begin{align}
    \langle\force',\force''\rangle_\rho\coloneqq
    \int_{\Omega}\force'\ccdot \cons(\force'')dx. \label{eq:metric}
\end{align}
It is actually symmetric and nondegenerate because we consider the Newtonian fluids:
For such fluids, the constitutive equation is linear,
$\cons(\force)=\lambda(\tr\force)\bm{\mathsf{I}}+2\mu\force$, and we have $\force'\ccdot \cons(\force'')=\lambda(\tr\force'')\bm{\mathsf{I}}\ccdot \force'+2\mu\force'\ccdot \force''$ and $\bm{\mathsf{I}}\ccdot \force'=\tr\force'$, so $\langle\cdot,\cdot\rangle_\rho$ is symmetric. 
As we assume the viscosities are always positive, the inner product can also be shown to be nondegenerate~\footnote{$\langle\force,\force\rangle_\rho
=\int [2\mu \bar{\force}\ccdot\bar{\force}+\zeta(\tr\force)^2]dx>0$ with $\bar{\force}=\force-\frac{1}{n}(\tr\force)\bm{\mathsf{I}}$. }.
The induced norm $\lVert\force'\rVert_\rho\coloneqq\sqrt{\langle\force',\force'\rangle_\rho}$ provides the geometric expression of the EPR $\dot{\Sigma}=\lVert\force\rVert_\rho^2$. 
We finally provide the space of thermodynamic forces as $\mathcal{F}=\{\force'=(\mathsf{F}'_{ij})\mid\force'={\force'}{}^\mathsf{T}, 
\lVert\force'\rVert_\rho<\infty,
\partial_i^2\mathsf{F}'_{jj}+\partial_j^2\mathsf{F}'_{ii}=2\partial_i\partial_j\mathsf{F}'_{ij}\}$, where the last condition is necessary for the force to have a velocity field leading to it.

While we only deal with the linear constitutive equation in the following, we can consider other constitutive equations, such as the generalized Newtonian models, to treat non-Newtonian fluids~\cite{bird1987dynamics}. Though Eq.~\eqref{eq:metric} is no longer symmetric then, it is expected that we can reproduce our results presented below based on some convex structure, as in chemical thermodynamics~\cite{kobayashi2022hessian,kolchinsky2022information}. 

\textit{Stochastic thermodynamics}.---%
Let us quickly review some results of stochastic thermodynamics whose generalizations we will consider. 
Stochastic thermodynamics studies thermally fluctuating, hence mesoscopic systems~\cite{seifert2012stochastic}. 
However, its framework has a deterministic flavor. 
Notably, the averaged EPR is given as the product of currents and thermodynamic forces, like classical macroscopic systems~\cite{de1962non}. 
This has enabled us to study chemical reaction networks (CRNs) in a stochastic-thermodynamic manner~\cite{rao2016nonequilibrium,yoshimura2021thermodynamic}. 

EPR decomposition~\cite{oono1998steady,hatano2001steady,esposito2010three,rao2016nonequilibrium,ge2016nonequilibrium,maes2014nonequilibrium,dechant2022geometric1,dechant2022geometric2,yoshimura2023housekeeping,kolchinsky2022information,kobayashi2022hessian} and thermodynamic uncertainty relations (TURs)~\cite{barato2015thermodynamic,pietzonka2016universal,dechant2018multidimensional,liu2020thermodynamic,horowitz2020thermodynamic,yoshimura2021thermodynamic,li2019quantifying,otsubo2020estimating,manikandan2020inferring} are crucial results of stochastic thermodynamics, which can also hold in CRNs. 
EPR decomposition breaks the total EPR, which quantifies the total dissipation in the whole system, into partial contributions, one required to keep the system out of equilibrium, called the housekeeping EPR, and the remainder, the excess EPR~\cite{oono1998steady}.
The excess EPR expresses the dissipation incurred by relaxation to a steady state~\cite{hatano2001steady} or quantifies the minimum dissipation to reproduce the dynamics~\cite{maes2014nonequilibrium}. 
What is significant about them is that they strengthen universal trade-off relations by capturing essentially distinct aspects of the dynamics. 

The thermodynamic uncertainty relation is an outstanding example of such universal relations and is one of the most meaningful findings of stochastic thermodynamics. 
It is usually formulated as a lower bound on the EPR that is inversely proportional to fluctuation measures~\cite{barato2015thermodynamic}. It reveals a trade-off that we must magnify fluctuations to reduce dissipation, or alternatively, larger dissipation is inevitable to get more accuracy~\cite{horowitz2020thermodynamic}. 
As the decomposed EPRs are smArxivaller than the total EPR, they can provide more strict bounds~\cite{shiraishi2018speed,dechant2022geometric1,dechant2022geometric2,yoshimura2023housekeeping,kolchinsky2022information,tuan2020unified}.

To generalize these results, we employ the so-called Maes--Neto\v{c}n\'{y} (MN) decomposition of EPR~\cite{maes2014nonequilibrium}, which utilizes the geometric structure of the thermodynamic forces~\cite{dechant2022geometric1,dechant2022geometric2,yoshimura2023housekeeping,kolchinsky2022information,kobayashi2022hessian}. 
It defines the housekeeping EPR as the squared distance (or divergence) between the current state and the subspace of conservative forces. 
A conservative force includes no ``cyclic'' contributions, leading the system to equilibrium. 
Therefore, the housekeeping EPR evaluates how the detailed balance is broken in the system. 
On the other hand, the excess EPR will be given as the minimum EPR to induce the dynamics. 
The connection between the two notions, the breaking of the detailed balance and the minimum dissipation, manifests itself in an orthogonal relation between forces. 

\textit{Result 1.\ EPR decomposition}.---%
As geometry has already been formulated, we now define the conservative subspace to generalize the MN decomposition to hydrodynamic systems. 
We define the conservative subspace $\mathcal{C}$ in the total force space $\mathcal{F}$ as the space of forces given by velocity fields that vanish on the boundary, $\mathcal{C}\coloneqq\{\force'\in\mathcal{F}\mid\exists\bm{u},\;\text{s.t.}\;\bm{u}|_{\partial\Omega}=\bm{0}\;\text{and}\;\force'=-\nabla^\mathsf{S}\bm{u}\}$, where $\cdot|_{\partial\Omega}$ means restriction onto $\partial\Omega$. 
If the thermodynamic force of a system (or the system, in short) is conservative, $\force\in\mathcal{C}$, then the system can be seen to be physically equivalent to a system on which no stress is exerted on the boundary, so we can expect it to relax to equilibrium globally.
It is noteworthy that the definition of $\mathcal{C}$ also means that when two forces $\force'$ and $\force''$ differ by an element in $\mathcal{C}$ ($\force'-\force''\in\mathcal{C}$), they will be given by velocity fields that share the boundary values.

The orthogonal complement of $\mathcal{C}$ is shown to be given as $\mathcal{C}_\rho^\perp=\{\force'\in\mathcal{F}\mid \nabla\cdot\cons(\force')=\bm{0}\}$. Let $\force''\in\mathcal{C}$. Then, there is $\bm{u}$ that vanishes on the boundary and satisfies $\force''=-\nabla^\mathsf{S}\bm{u}$.
Noting the symmetry $\cons(\force)^\mathsf{T}=\cons(\force)$, we see that integration by parts leads to
\begin{align*}
    \langle\force'',\force'\rangle_\rho
    =\int_\Omega\bm{u}\cdot(\nabla\cdot\cons(\force'))dx,
\end{align*}
which implies $\nabla\cdot\cons(\force')$ should be zero everywhere. 
As the inner product is nondegenerate, we can decompose the thermodynamic force $\force\in\mathcal{F}$ into the conservative part $\force_\mathrm{c}\in\mathcal{C}$ and the nonconservative part $\force_\mathrm{nc}\in\mathcal{C}_\rho^\perp$ as $\force=\force_\mathrm{c}+\force_\mathrm{nc}$ with the orthogonal relation $\langle\force_\mathrm{c},\force_\mathrm{nc}\rangle_\rho=0$.

To discuss the physical meaning of $\mathcal{C}_\rho^\perp$, consider two systems with the same reversible current $\revcurr$ and different thermodynamic forces $\force'$ and $\force''$. 
If the difference is in $\mathcal{C}_\rho^\perp$, we get $\nabla\cdot\cons(\force')=\nabla\cdot\cons(\force'')$ due to the linearity of $\cons(\cdot)$. 
Hence, the dynamics they yield will coincide in terms of momentum, as the irreversible term $\nabla\cdot\irrcurr$ in the continuity equation is given by $\nabla\cdot\cons(\force')=\nabla\cdot\cons(\force'')$. 
Since $\mathcal{C}_\rho^\perp$ contains the zero tensor field, it can be understood as the space of forces that do not affect the dynamics. 
These interpretations implicitly assume that the momentum and the thermodynamic force can be separately considered, which may not always be true in a real hydrodynamic situation. However, we can expect that they would give a dynamical characterization of the optimal transport theory of vector fields like the Benemou--Brenier formula~\cite{benamou2000computational}.

Now that the geometry and the conservative subspace are obtained, we can define an EPR decomposition for hydrodynamic systems.
Following the MN prescription, we decompose the EPR as follows: First, the housekeeping EPR is defined as
\begin{align}
    \dot{\Sigma}^\mathrm{hk}
    \coloneqq\min_{\force'\in\mathcal{C}}\,\lVert\force-\force'\rVert_\rho^2.
    \label{eq:hkdef}
\end{align}
Measuring the distance between the present state $\force$ and the conservative subspace $\mathcal{C}$, the housekeeping EPR quantifies the dissipation stemming from the boundary motion that makes the system out of equilibrium. 
Next, we define the excess EPR as
\begin{align}
    \dot{\Sigma}^\mathrm{ex}
    \coloneqq\min_{\force'\in\mathcal{F}}\lVert\force'\rVert_\rho^2\quad\mathrm{s.t.}\quad\nabla\cdot\cons(\force')=\nabla\cdot\cons(\force).
    \label{eq:exdef}
\end{align}
The condition can be rephrased by $\force'-\force\in\mathcal{C}_\rho^\perp$, which leads to another expression $\dot{\Sigma}^\mathrm{ex}=\min_{\force'\in\mathcal{C}_\rho^\perp}\lVert\force-\force'\rVert_\rho^2$. 
The definition shows that the excess EPR is the minimum dissipation to induce the same dynamics as the original system in terms of irreversible currents. 
They can be shown to sum up to the total EPR, $\dot{\Sigma}=\dot{\Sigma}^\mathrm{hk}+\dot{\Sigma}^\mathrm{ex}$, and to be given by $\force_{\mathrm{c}}$ and $\force_{\mathrm{nc}}$ as $\dot{\Sigma}^\mathrm{hk}=\lVert\force_{\mathrm{nc}}\rVert_\rho^2$ and $\dot{\Sigma}^\mathrm{ex}=\lVert\force_{\mathrm{c}}\rVert_\rho^2$. 
As the proof of these results only requires slight modifications of the existing proof of the MN decomposition, it is provided in Supplementary Material~\cite{smArxiv}.

\textit{Result 2.\ TUR.---}
In addition to the definitions using minimization, the EPRs also have maximization expressions, which we are going to see finally yield a TUR. 
First, we give the maximization representation of the total EPR
\begin{align}
    \dot{\Sigma}=\max_{\force'\in\mathcal{F}}\frac{(\langle\force',\force\rangle_\rho)^2}{\lVert\force'\rVert_\rho^2}. \label{eq:tur}
\end{align}
This is derived from the Cauchy--Schwarz inequality $\lVert\force'\rVert_\rho^2\lVert\force''\rVert_\rho^2\geq(\langle\force',\force''\rangle_\rho)^2$ and setting $\force''=\force$. 
The equality is actually achieved when $\force'=\alpha\force$ ($\alpha\in\mathbb{R}$). 
We can also provide the housekeeping and excess EPRs with the maximization expressions
\begin{align}
    \dot{\Sigma}^\mathrm{hk}
    =\max_{\force'\in\mathcal{C}_\rho^\perp}\frac{(\langle\force',\force\rangle_\rho)^2}{\lVert\force'\rVert_\rho^2},\quad
    \dot{\Sigma}^\mathrm{ex}
    =\max_{\force'\in\mathcal{C}}\frac{(\langle\force',\force\rangle_\rho)^2}{\lVert\force'\rVert_\rho^2}.
    \label{eq:tur2}
\end{align}
Again, these are derived from the Cauchy--Schwarz inequality: Choose $\force''=\force_\mathrm{nc}$. Then, as long as $\force'\in\mathcal{C}_\rho^\perp$, we have $\langle\force',\force_\mathrm{nc}\rangle_\rho=\langle\force',\force\rangle_\rho$ because $\force-\force_\mathrm{nc}\in\mathcal{C}$. Since the equality holds when $\force'=\alpha\force_\mathrm{nc}$ ($\alpha\in\mathbb{R}$), we get the formula for the housekeeping EPR. The excess version can be proved similarly.

These expressions lead to a TUR similar to the short-time TUR derived in stochastic thermodynamics~\cite{otsubo2020estimating}. 
Consider a time dependent vector field $\bm{a}=\bm{a}(\bm{x},t)$ such that $\nabla^{\mathsf{S}}\bm{a}\in\mathcal{C}$ and $\nabla\cdot\bm{a}=0$ for all $t$. 
The second equality in Eq.~\eqref{eq:tur2} implies $\dot{\Sigma}^\mathrm{ex}\geq (\langle\nabla^\mathsf{S}\bm{a},\force\rangle_\rho)^2/\lVert\nabla^\mathsf{S}\bm{a}\rVert_\rho^2$ for such $\bm{a}$. 
Further calculation, presented in~\cite{smArxiv}, finally leads to the TUR
\begin{align}
    \dot{\Sigma}^\mathrm{ex}\geq \frac{(\dot{J}_{\bm{a}})^2}{\mu_\mathrm{max}\lVert\nabla\bm{a}\rVert^2}. \label{eq:extur}
\end{align}
Here, $\dot{J}_{\bm{a}}\coloneqq\frac{d}{dt}\int_\Omega\rho\bm{v}\cdot\bm{a}dx-\int_{\Omega}\rho\bm{v}\cdot\frac{D\bm{a}}{Dt}dx$ gives the changing rate of the quantity $\int_\Omega\rho\bm{v}\cdot\bm{a}dx$ minus the local effect of $\bm{a}$'s dynamics (so it only reflects the system's dynamics). 
The interpretation of $\dot{J}_{\bm{a}}$ as a changing rate gets clearer 
if we further assume that $\bm{a}$ is given by a time-independent vector field $\bm{A}(\bm{x})$ and the pathline $\bm{\phi}_t(\bm{x})$ as $\bm{a}(\bm{x},t)=\bm{A}(\bm{\phi}_t^{-1}(\bm{x}))$, 
because then $\frac{D\bm{a}}{Dt}=\bm{0}$ holds and $\dot{J}_{\bm{a}}$ becomes the time derivative $\frac{d}{dt}\int_{\Omega}\rho\bm{v}\cdot\bm{a}dx$. 
On the other hand, the denominator consists of quantities that represent fluctuations: the maximum shear viscosity $\mu_\mathrm{max}=\max_{\bm{x}\in\Omega}\mu(\rho(\bm{x}))$ and the squared norm of tensor $\nabla\bm{a}$, $\lVert\nabla\bm{a}\rVert\coloneqq\sqrt{\int_\Omega\sum_{i,j}(\partial_ia_j)^2dx}$. 
The former can be calculated from microscopic hydrodynamic fluctuations by the Green--Kubo relation~\cite{sasa2014derivation}, though the fluctuations are not considered in this Letter. 
The latter obviously represents the spatial fluctuations of the vector field of interest $\bm{a}$. 
Therefore, the inequality in Eq.~\eqref{eq:extur} gives a lower bound on the excess (hence, the total) EPR by the ratio between a rate and fluctuations, 
and it can be seen as a kind of TUR.

\begin{figure}
    \centering
    \includegraphics[width=\linewidth]{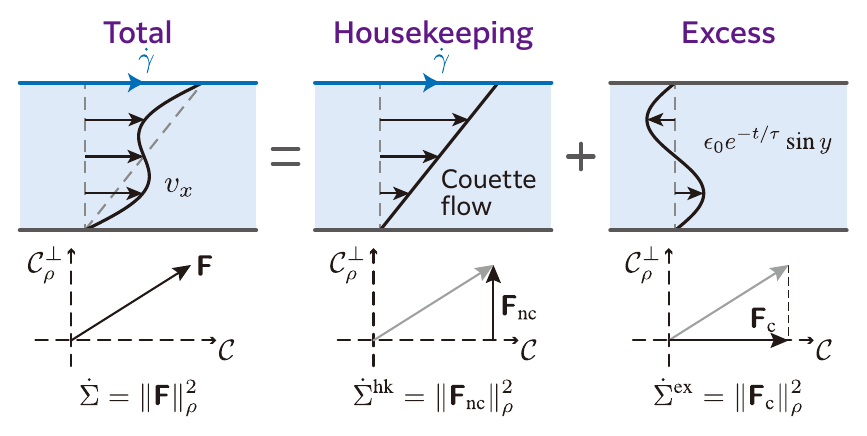}
    \caption{Decomposition of the EPR in the perturbed Couette flow. We consider a fluid exerted shear on by a wall moving at speed $\dot{\gamma}$. The sinusoidal perturbation will vanish and the steady Couette flow will remain. In our geometric decomposition, the housekeeping EPR represents the dissipation incurred by the moving wall and is given by the Couette flow. The excess part, on the other hand, stems from the transient mode, vanishing in the relaxation. Note that although the steady flow emerges in the decomposition, the decomposition can be implemented solely with the instantaneous fields and the boundary condition, without referring to the steady state. }
    \label{fig:couette}
\end{figure}

\textit{Example.---}
We exemplify the decomposition through a toy model of shear flow in $\Omega=S^1\times[0,1]$ ($S^1$ is the 1-sphere of length 1. See Fig.~\ref{fig:couette}). 
The TUR is also discussed by this model in Supplementary Material~\cite{smArxiv}. 
The boundary is composed of the top and the bottom line, $\partial\Omega=S^1\times(\{0\}\cup\{1\})$. 
The bottom $S^1\times\{0\}$ is fixed while the top $S^1\times\{1\}$ is moved in the $x$ direction at rate $\dot{\gamma}$. 
We assume the no-slip boundary condition, i.e., every velocity field should satisfy $\bm{v}(x,0)=(0,0)$ and $\bm{v}(x,1)=(\dot{\gamma},0)$.
Let the initial density and pressure be uniform and the initial velocity field be the Couette flow $(\dot{\gamma}y,0)$ plus the perturbation $(\epsilon_0\sin(2\pi y),0)$. 
The ansatz $\rho(x,y,t)=\rho\,(\mathrm{const.})$, $p(x,y,t)=\mathrm{const.}$, and $\bm{v}(x,y,t)=(\dot{\gamma}y+\epsilon(t)\sin(2\pi y),0)$ solve the Navier--Stokes equation with $\epsilon(t)=\epsilon_0e^{-t/\tau}$ ($\tau=\rho/(4\pi^2\mu)$). 
The thermodynamic force then reads $\mathsf{F}_{xy}=\mathsf{F}_{yx}=-[\dot{\gamma}/2+\pi\epsilon\cos(2\pi y)]$ and $\mathsf{F}_{xx}=\mathsf{F}_{yy}=0$, and
the housekeeping EPR is provided by
\begin{align*}
    \dot{\Sigma}^\mathrm{hk}
    =&\min_{\bm{u}}
    \int_0^1dy\int_{S^1}dx\,\\
    &\Big[2\mu\big\{(\partial_xu_x)^2+(\partial_yu_y)^2\big\}
    +\lambda(\partial_xu_x+\partial_yu_y)^2\\
    &\phantom{tea}+\mu\big\{\partial_xu_y+\partial_yu_x-\dot{\gamma}-2\pi\epsilon\cos(2\pi y)\big\}^2\Big]
\end{align*}
with condition $\bm{u}|_{\partial\Omega}=\bm{0}$. 
It is solved by $\bm{u}=(\epsilon\sin(2\pi y),0)$. Note the optimization can be done regardless of the time dependence. 
Finally, we get the explicit decomposition as
\begin{align}
    \dot{\Sigma}^\mathrm{hk}=\mu\dot{\gamma}^2,\quad
    \dot{\Sigma}^\mathrm{ex}= 2\pi^2\epsilon_0^2\mu e^{-2t/\tau}.
\end{align}
In this model, the housekeeping term reflects the breaking of detailed balance by the shear, which causes the steady-state dissipation $\mu\dot{\gamma}^2$, while the excess term gives the extra dissipation due to the relaxation with magnitude quadratically proportional to the perturbation strength $\epsilon_0$. 
Therefore, now the decomposition is consistent with the original idea by Oono and Paniconi presented in Ref.~\cite{oono1998steady}. 
However, we should be aware that such an understanding will not be valid if the system does not relax to a Stokes flow, which satisfies $\nabla\cdot\bm{v}=0$ and $\nabla\cdot\cons(\force)=\nabla p$ (in Supplementary Material~\cite{smArxiv}, we discuss this point in more depth). 
Nonetheless, the EPR decomposition is always feasible, keeping its own physical meaning, even if there are no steady states because it only needs the instantaneous fields and the boundary conditions. 

\textit{Conclusion}.---%
In summary, we have reformulated the thermodynamic theory of classical hydrodynamics so that it adapts to stochastic thermodynamics and derived the EPR decomposition and the TUR in hydrodynamics. 
We regard minus the symmetrized gradient of a velocity field, rather than just the gradient, as the thermodynamic force to obtain a formalismArxiv similar to stochastic and chemical thermodynamics. 
The geometry of thermodynamic forces allowed us to generalize the two significant results of stochastic thermodynamics. 
We showed that the MN-type decomposition actually works in hydrodynamics and EPRs have a TUR-type lower bound, which is given by rate and fluctuation.

These results can be more meaningful if what ``minimum dissipation'' implies is clearer or if a nice choice of $\bm{A}$ in the TUR is found. 
As noted, the excess EPR formulated as minimum dissipation will suggest a way to extend optimal transport theory~\cite{villani2009optimal,nakazato2021geometrical}. 
We could expect that focusing on particles' motion in a fluid would be relevant for the choice of $\bm{A}$. 

Another promising direction is to remove the assumption of Newtonianity. 
It so far ensures the linear relation between the thermodynamic force $\force$ and the irreversible current $\irrcurr$, while excluding non-Newtonian systems, such as polymer solutions~\cite{bird1987dynamics}.  
On the other hand, in stochastic thermodynamics, it has been proved that information geometry (also known as the Hessian geometry) enables one to deal with nonlinear relation between thermodynamic forces and currents~\cite{kobayashi2022hessian,kolchinsky2022information}. 
As the current framework is established only relying on the existence of the constitutive equation $\bm{\Pi}_\rho$ rather than the linearity, we believe that our results will be naturally extended to non-Newtonian systems. 

\begin{acknowledgments}
K.Y.\ and S.I.\ thank Artemy Kolchinsky, Ken Hiura, Ryuna Nagayama, and Naruo Ohga for their suggestive comments, and Shin-ichi Sasa for fruitful discussion. K.Y.\ thanks Kazumasa A.\ Takeuchi for giving reasonable comments regularly. 
K.Y.\ is supported by Grant-in-Aid for JSPS Fellows (Grant No.~22J21619). 
S.I. is supported by JSPS KAKENHI Grants No.~19H05796, No.~21H01560, No.~22H01141, No.~23H00467, and No.~24H00834, 
JST ERATO Grant No.~JPMJER2302, 
and UTEC-UTokyo FSI Research Grant Program. 
\end{acknowledgments}

%

\clearpage
\renewcommand{\theequation}{S\arabic{equation}}
\renewcommand{\thetable}{S\arabic{table}}
\onecolumngrid
\begin{center}
    \large\textbf{Supplemental Material for}\\
    \textbf{``Two applications of stochastic thermodynamic to hydrodynamics''}\\
    \medskip
    {\small Kohei Yoshimura and Sosuke Ito}
\end{center}
\section{On the incompressible case}
In this section, we consider incompressible systems, where the incompressibility $\nabla\cdot\bm{v}=0$ with homogeneity $\rho=\mathrm{const.}$ is assumed. 
These assumptions only lead to a few modifications of the definitions of force spaces. 

First of all, it becomes necessary to assume thermodynamic forces to be traceless because $\tr\force=-\sum_{i}\partial_iv_i=-\nabla\cdot\bm{v}$. 
Thus, the space of forces should be replaced with $\mathring{\mathcal{F}}=\{\force'\in\mathcal{F}\mid\tr\force'=0\}$, where we suppose $\mathring{}$ to indicate the elements of the set are traceless. The constitutive equation always reads $\cons(\force)=2\mu\force$ and $\lambda$ will not appear. 
The conservative subspace is simply given as $\mathring{\mathcal{C}}=\mathcal{C}\cap\mathring{\mathcal{F}}$. 
On the other hand, its orthogonal complement requires further consideration.
When $\force''\in\mathring{\mathcal{C}}$, it is given as $\force''=-\nabla^\mathsf{S}\bm{u}$ by $\bm{u}$ such that $\bm{u}|_{\partial\Omega}=\bm{0}$ and $\nabla\cdot\bm{u}=0$. Thus, 
\begin{align}
    \langle\force',\force''\rangle_\rho
    =\int_{\Omega}\bm{u}\cdot(\nabla\cdot\cons(\force'))dx=0
\end{align}
means that $\nabla\cdot\cons(\force')$ is given as the gradient $\nabla p$ of a scalar function $p$ rather than $\nabla\cdot\cons(\force')=\bm{0}$. 
Therefore, the orthogonal complement becomes $\mathring{\mathcal{C}}_\rho^\perp=\{\force'\in\mathring{\mathcal{C}}\mid \exists p,\;\nabla\cdot\cons(\force')=-\nabla p\}$. 
As is obvious from the proof given later, the EPR decomposition is independent of the details of the subspaces, so it also works for incompressible systems virtually as it is. The only difference is that we need to use the expression $\dot{\Sigma}^\mathrm{ex}=\min_{\force'\in\mathring{\mathcal{C}}_\rho^\perp}\lVert\force-\force'\rVert_\rho^2$ presented below Eq.~(5) in the main text to define the excess EPR, and the condition $\nabla\cdot\cons(\force')=\nabla\cdot\cons(\force)$ in Eq.~(5) is inappropriate for incompressible systems. 

The orthogonal complement turns out to be related to the Stokes equation. 
Since $\force'\in\mathring{\mathcal{C}}$ is provided by incompressible field $\bm{v}$, we have $\nabla\cdot\cons(\force')=-\mu\Delta \bm{v}$, where $\Delta=\nabla\cdot\nabla$. 
Therefore, the condition $\force=-\nabla^\mathsf{S}\bm{v}\in\mathring{\mathcal{C}}_\rho^\perp$ means $\mu\Delta\bm{v}-\nabla p=\bm{0}$, which is the Stokes equation~\cite{batchelor1967introduction}. 
This fact connects the decomposition to the original philosophy of EPR decomposition when we consider incompressible systems governed by the nonstationary Stokes equation, $\rho\partial_t\bm{v}=-\nabla p+\mu\Delta\bm{v}$. 
Originally, the housekeeping EPR was proposed to give the dissipation in a nonequilibrium steady state~\cite{oono1998steady}. 
Because the nonconservative force $\force_\mathrm{nc}$, which determines the housekeeping EPR, belongs to the orthogonal complement, it is actually stationary in the sense that the corresponding velocity field(s) becomes a stationary solution of the nonstationary Stokes equation. 
Thus, the housekeeping EPR quantifies the steady-state dissipation. 
On the other hand, in the general (Navier--Stokes or incompressible) case, the subspaces are not associated with the genuine stationarity (but only with the condition $\nabla\cdot\irrcurr=\bm{0}$). 

It is worth noting that the housekeeping EPR in the incompressible setup is connected to the Helmholtz minimum dissipation theorem~\cite{helmholtz1868theorie,batchelor1967introduction}. 
The theorem states that, for a given boundary condition of velocity fields, the minimum EPR is provided by the solution of the Stokes equation. 
We can show that the housekeeping EPR is nothing but the minimum dissipation corresponding to the boundary velocity field at the moment $\bm{v}(t)|_{\partial\Omega}$. 
First, note that when the difference between two forces belongs to $\mathring{\mathcal{C}}$, they can be induced by velocity fields that share the boundary values. 
Thus, because $\force_\mathrm{nc}-\force=\force_\mathrm{c}\in\mathring{\mathcal{C}}$, the thermodynamic force and its nonconservative component share the boundary condition. 
Moreover, $\force_\mathrm{nc}\in\mathring{\mathcal{C}}_\rho^\perp$ means that the velocity fields that lead to $\force_\mathrm{nc}$ satisfies the Stokes equation, as discussed above. 
Therefore, the housekeeping EPR $\lVert\force_\mathrm{nc}\rVert_\rho^2$ is incurred by a velocity field that coincides with $\bm{v}(t)$ on the boundary and satisfies the Stokes equation $\mu\Delta\bm{v}'=\nabla p$ and is the minimum dissipation in the sense of the Helmholtz minimum dissipation theorem. 

\section{Derivation of the decomposition}
Let us show the decomposition
\begin{align}
    \dot{\Sigma}=\dot{\Sigma}^\mathrm{hk}+\dot{\Sigma}^\mathrm{ex},
\end{align}
with $\dot{\Sigma}=\lVert\force\rVert_\rho^2$, $\dot{\Sigma}^\mathrm{hk}=\min_{\force'\in\mathcal{C}}\lVert\force-\force'\rVert_\rho^2$, and $\dot{\Sigma}^\mathrm{ex}=\min_{\force'\in\mathcal{C}_\rho^\perp}\lVert\force-\force'\rVert_\rho^2$. 
Though it is just the Hilbert projection theorem~\cite{rudin1987real}, we provide a proof by assuming a part of the theorem; we assume there is $\force''\in\mathcal{C}$ such that $\force-\force''\in\mathcal{C}_\rho^\perp$. 
We also define $\force_\mathrm{c}=\mathop{\mathrm{arg\,min}}_{\force'\in\mathcal{C}}\lVert\force-\force'\rVert_\rho^2$. 
Then, we have
\begin{align}
    \lVert\force-\force_\mathrm{c}\rVert_\rho^2
    &=\lVert\force-\force''+\force''-\force_\mathrm{c}\rVert_\rho^2
    =\lVert\force-\force''\rVert_\rho^2
    +\lVert\force''-\force_\mathrm{c}\rVert_\rho^2
    +2\langle\force-\force'',\force''-\force_\mathrm{c}\rangle_\rho\\
    &=\lVert\force-\force''\rVert_\rho^2
    +\lVert\force''-\force_\mathrm{c}\rVert_\rho^2
    \geq \lVert\force-\force''\rVert_\rho^2,
\end{align}
where the third term in the rightmost equation in the first line vanishes since $\force-\force''\in\mathcal{C}_\rho^\perp$ and $\force''-\force_\mathrm{c}\in\mathcal{C}$. Therefore, due to the definition of $\force_\mathrm{c}$, we see that $\force''$ must be $\force_\mathrm{c}$ and $\force_\mathrm{nc}\coloneqq\force-\force_\mathrm{c}$ must be in $\mathcal{C}_\rho^\perp$ (otherwise, the difference $\lVert\force''-\force_\mathrm{c}\rVert_\rho^2$ ruins the minimizing property of $\force_\mathrm{c}$). 
In addition, we get the explicit expression of the housekeeping EPR $\dot{\Sigma}^\mathrm{hk}=\lVert\force_\mathrm{nc}\rVert_\rho^2$. 

Similarly we can define $\force_\mathrm{nc}'\coloneqq\mathop{\mathrm{arg\,min}}_{\force'\in\mathcal{C}_\rho^\perp}\lVert\force-\force'\rVert_\rho^2$ to derive $\force_\mathrm{c}'\coloneqq\force-\force_\mathrm{nc}'\in\mathcal{C}$ and $\dot{\Sigma}^\mathrm{ex}=\lVert\force_\mathrm{c}'\rVert_\rho^2$. 
Now that we need to prove $\force_\mathrm{nc}=\force_\mathrm{nc}'$, in other words, the uniqueness of the decomposition of force. 
Considering the definiteness of the norm, the proof is straightforward. 
Because $\force=\force_\mathrm{c}+\force_\mathrm{nc}=\force_\mathrm{c}'+\force_\mathrm{nc}'$, 
\begin{align}
    0&=\lVert\force_\mathrm{c}+\force_\mathrm{nc}-(\force_\mathrm{c}'+\force_\mathrm{nc}')\rVert_\rho^2
    =\lVert\force_\mathrm{c}-\force_\mathrm{c}'+\force_\mathrm{nc}-\force_\mathrm{nc}'\rVert_\rho^2\\
    &=\lVert\force_\mathrm{c}-\force_\mathrm{c}'\rVert_\rho^2
    +\lVert\force_\mathrm{nc}-\force_\mathrm{nc}'\rVert_\rho^2
    +2\langle\force_\mathrm{c}-\force_\mathrm{c}',\force_\mathrm{nc}-\force_\mathrm{nc}'\rangle_\rho\\
    &=\lVert\force_\mathrm{c}-\force_\mathrm{c}'\rVert_\rho^2
    +\lVert\force_\mathrm{nc}-\force_\mathrm{nc}'\rVert_\rho^2,
\end{align}
where the inner product in the second line vanishes due to the orthogonality. 
Since both terms in the last line must be zero, the decomposition of force is unique. 
Consequently, we finally obtain $\Sigma^\mathrm{hk}=\lVert\force_\mathrm{nc}\rVert_\rho^2$ and $\Sigma^\mathrm{ex}=\lVert\force_\mathrm{c}\rVert_\rho^2$, and the decomposition $\dot{\Sigma}=\dot{\Sigma}^\mathrm{hk}+\dot{\Sigma}^\mathrm{ex}$ is evident, given  $\langle\force_\mathrm{c},\force_\mathrm{nc}\rangle_\rho=0$. 

\section{Derivation of the TUR}
We give a proof of the TUR
\begin{align}
    \dot{\Sigma}^\mathrm{ex}\geq \frac{(\dot{J}_{\bm{a}})^2}{\mu_{\mathrm{max}}\lVert\nabla\bm{a}\rVert^2}, 
\end{align}
which was presented as Eq.~(8) in the main text. 
The quantities on the right-hand side are defined as $\dot{J}_{\bm{a}}\coloneqq\frac{d}{dt}\int_\Omega\rho\bm{v}\cdot\bm{a}dx-\int_{\Omega}\rho\bm{v}\cdot\frac{D\bm{a}}{Dt}dx$, $\mu_\mathrm{max}=\max_{\bm{x}\in\Omega}\mu(\rho(\bm{x}))$, and $\lVert\nabla\bm{a}\rVert\coloneqq\sqrt{\int_\Omega\sum_{i,j}(\partial_ia_j)^2dx}$. Vector field $\bm{a}$ is assumed to be incompressible $\nabla\cdot\bm{a}=0$ and its symmetrized gradient to belong to $\mathcal{C}$ (namely, it is physically equivalent to a velocity field that vanishes on the boundary).

To show the relation, we first check the following equations:
\begin{align}
    \langle\nabla^\mathsf{S}\bm{a},\force\rangle_\rho
    =\frac{d}{dt}\int_{\Omega}\rho\bm{a}\cdot\bm{v}\,dx
    -\int_{\Omega}\rho\bm{v}\cdot\frac{D\bm{a}}{Dt}\,dx,\quad
    \lVert\nabla^\mathsf{S}\bm{a}\rVert_\rho^2
    \leq\mu_{\mathrm{max}}\lVert\nabla\bm{a}\rVert^2. \label{eq:1and2}
\end{align}
Let us confirm the first one. Because of the Navier--Stokes equation $\rho\frac{D\bm{v}}{Dt}=-\nabla p-\nabla\cdot\cons(\force)$, 
\begin{align}
    \langle\nabla^\mathsf{S}\bm{a},\force\rangle_\rho
    &=\int_{\Omega}\nabla\bm{a}\ccdot\cons(\force)\,dx
    =-\int_{\Omega}\bm{a}\cdot(\nabla\cdot\cons(\force))\,dx\\
    &=\int_\Omega\bm{a}\cdot\left(\rho\frac{D\bm{v}}{Dt}+\nabla p\right)\,dx
    =\int_\Omega\rho\bm{a}\cdot\frac{D\bm{v}}{Dt}\,dx,
    \label{eq:1-1}
\end{align}
where we did integration by parts and derived the last equation using the properties of $\bm{a}$, $\bm{a}|_{\partial\Omega}=\bm{0}$ and $\nabla\cdot\bm{a}=0$. 
Applying the identity~\cite{batchelor1967introduction}
\begin{align}
    \frac{d}{dt}\int_\Omega \rho\phi\,dx
    =\int_\Omega \rho\frac{D\phi}{Dt}\,dx
\end{align}
to $\phi=\bm{a}\cdot\bm{v}$ shows
\begin{align}
    \frac{d}{dt}\int_{\Omega}\rho\bm{a}\cdot\bm{v}\,dx
    =\int_{\Omega}\Big(\rho\bm{a}\cdot\frac{D\bm{v}}{Dt}
    +\rho\bm{v}\cdot\frac{D\bm{a}}{Dt}\Big)\,dx. \label{eq:1-2}
\end{align}
Combining Eqs.~\eqref{eq:1-1} and \eqref{eq:1-2} leads to the first equality of Eq.~\eqref{eq:1and2}. 

The inequality in Eq.~\eqref{eq:1and2} is proved as follows:
\begin{align}
    \lVert\nabla^\mathsf{S}\bm{a}\rVert_\rho^2
    &=2\int_{\Omega}\mu(\rho)\nabla\bm{a}\ccdot\nabla^\mathsf{S}\bm{a}dx
    \leq 2\mu_{\mathrm{max}}\int_{\Omega}\nabla\bm{a}\ccdot\nabla^\mathsf{S}\bm{a}dx\\
    &=\mu_{\mathrm{max}}\left[\int_{\Omega}\nabla\bm{a}\ccdot\nabla\bm{a}dx+\int_{\Omega}\nabla\bm{a}\ccdot(\nabla\bm{a})^\mathsf{T}dx\right]\\
    &=\mu_{\mathrm{max}}\left[\int_{\Omega}\nabla\bm{a}\ccdot\nabla\bm{a}dx-\int_{\Omega}\nabla(\nabla\cdot\bm{a})\cdot\bm{a}dx\right]\\
    &=\mu_{\mathrm{max}}\int_{\Omega}\nabla\bm{a}\ccdot\nabla\bm{a}dx,
\end{align}
where we used $\nabla\cdot\bm{a}=0$ to derive $\cons(\nabla^\mathsf{S}\bm{a})=2\mu\nabla^\mathsf{S}\bm{a}$ in the first line and to obtain the last equality, and used the fact $\bm{a}\in\mathcal{C}$ to do integration by parts to get the third line. 

We finally apply Eq.~\eqref{eq:1and2} to the maximization formula provided in Eq.~(7) to obtain
\begin{align}
    \dot{\Sigma}^\mathrm{ex}
    \geq \frac{(\langle\nabla^\mathsf{S}\bm{a},\force\rangle_\rho)^2}{\lVert\nabla^\mathsf{S}\bm{a}\rVert_\rho^2}
    \geq \frac{(\dot{J}_{\bm{a}})^2}{\mu_{\mathrm{max}}\lVert\nabla\bm{a}\rVert^2}. 
\end{align}

\section{Example of the TUR}
Let us discuss the TUR by the Couette flow presented in the main text. 
The velocity field $\bm{v}(x,y,t)=(\dot{\gamma}y+\epsilon(t)\sin(2\pi y),0)$ with $\epsilon(t)=\epsilon_0e^{-t/\tau}$ and $\tau=\rho/(4\pi^2\mu)$ has the pathline 
\begin{align}
    \bm{\phi}_t(x,y)
    =\left[x+\dot{\gamma}yt+\tau\epsilon_0\sin(2\pi y)(1-e^{-t/\tau}),y\right],
\end{align}
whose inverse is given as 
\begin{align}
    \bm{\phi}_t^{-1}(X,Y)
    =\left[X-\dot{\gamma}Yt-\tau\epsilon_0\sin(2\pi Y)(1-e^{-t/\tau}),Y\right]. 
\end{align}
The incompressible condition $\nabla\cdot\bm{a}=0$ ($\bm{a}(\bm{X},t)=\bm{A}(\bm{\phi}_t^{-1}(\bm{X},t))$) now reads
\begin{align}
    0
    &=\partial_XA_x(\bm{\phi}_t^{-1}(\bm{X},t))
    +\partial_YA_y(\bm{\phi}_t^{-1}(\bm{X},t))\\
    &=\partial_xA_x(\bm{\phi}_t^{-1}(\bm{X},t))
    +\partial_xA_y(\bm{\phi}_t^{-1}(\bm{X},t))
    \left[-\dot{\gamma}t-2\pi\tau\epsilon_0\cos(2\pi Y)(1-e^{-t/\tau})\right]
    +\partial_yA_y(\bm{\phi}_t^{-1}(\bm{X},t)).
\end{align}
If we set $A_y(x,y)=f(y)$ with a differentiable function $f$, this condition is satisfied by $A_x(x,y)=-xf'(y)+g(y)$ with a function $g$. 
Since the system is periodic in the $x$-direction, $A_x(x+1,y)=A_x(x,y)$, which results in $f'(y)=0$. 
Finally, the boundary condition $\bm{A}(y=0)=\bm{A}(y=1)=\bm{0}$ requires the conditions $f(y)=0$ and $g(0)=g(1)=0$. 
Therefore, the TUR is valid for any vector field of the form $\bm{a}(x,y)=(g(y),0)$ with any $g$ such that $g(0)=g(1)=0$. 

Plugging this form of $\bm{a}$ into Eq.~(8) finally leads to the lower bound
\begin{align}
    \dot{\Sigma}^{\mathrm{ex}}
    =2\pi^2\epsilon_0^2\mu e^{-2t/\tau}
    \geq 
    \frac{\mu(4\pi^2\epsilon_0g^{(1)})^2}{\int_0^1(g'(y))^2dy}e^{-2t/\tau},
\end{align}
where $g^{(1)}=\int_0^1g(y)\sin(2\pi y)dy$. 
We can directly prove the inequality from the Cauchy--Schwarz inequality
\begin{align}
    \left(\int_0^1g'(y)\cos(2\pi y)dy\right)^2
    \leq \int_0^1 (g'(y))^2dy
    \int_0^1 \cos^2(2\pi y)dy
    =\frac{1}{2}\int_0^1 (g'(y))^2dy
\end{align}
because
\begin{align}
    g^{(1)}
    =\int_0^1g(y)\sin(2\pi y)dy
    =-\frac{1}{2\pi}\int_0^1g'(y)\cos(2\pi y)dy. 
\end{align}
Thus, the inequality is tight when $g'(y)$ approximates $\cos(2\pi y)$ well. 
For example, if we set $g(y)=y(y-1/2)(y-1)$, then $g^{(1)}=3/(4\pi^2)$ and $\int_0^1(g'(y))^2dy=1/20$, so the inequality relation between the prefactors of $e^{-2t/\tau}$ becomes $2\pi^2\epsilon_0^2\mu\geq (180/\pi^2)\epsilon_0^2\mu$. 
Since $(180/\pi^2)/2\pi^2=0.9239...$, we can confirm that the TUR provides a tight bound. 

\section{On Excess EPR in steady state}

The excess EPR vanishes in a steady state only if the steady state satisfies $\nabla\cdot\irrcurr=\bm{0}$ in compressible systems and if $\nabla\cdot\irrcurr=\nabla q$ with a scalar function $q$ in incompressible systems. 
These conditions can be seen as the generalizations of the Stokes equation. 
However, in general, they are not the case because the stationarity condition only requires $\nabla\cdot\curr=\bm{0}$, which does not mean $\dot{\Sigma}^\mathrm{ex}=0$. 
This is anomalous in light of the original philosophy of the housekeeping--excess decomposition that the excess EPR evaluates how far from the steady states the system is. 
Nonetheless, this anomaly has been observed in other kinds of systems. 
In underdamped Langevin systems, the decomposition is known not to work properly without an extra term that is hard to interpret physically~\cite{spinney2012entropy}.
The same issue exists in the decomposition in Markov jump processes with odd variables~\cite{spinney2012nonequilibrium}, where the Maes--Neto\v{c}n\'{y} excess EPR does not vanish in a steady state~\cite{kolchinsky2022information}.
The systems are summarized in Table~\ref{tab:comparison}: Theoretical frameworks grouped as Type 1 commonly neglect the inertia of the system, whereas those grouped as Type 2 respect it. 
Hence, we can see inertia affects the housekeeping--excess decomposition in a wide range of systems, ranging from mesoscopic to macroscopic systems. 

\begin{table}[h]
    \centering
    \begin{tabular}{c || l c l c l}
        & \textit{Langevin} && \textit{Markov jump} && \textit{Hydrodynamic }\\\hline\hline
        Type 1& Overdamped && w/o odd variables && Stokes eq.\\\hline
        Type 2& Underdamped && w/ odd variables && Navier--Stokes eq.\\\hline
    \end{tabular}
    \caption{In type-1 systems, the excess EPR vanishes in a steady state, while it does not, or has an unphysical additional term, in type-2 systems. }
    \label{tab:comparison}
\end{table}

\end{document}